# Optical parametric amplification of phase-stable terahertz-to-midinfrared pulses studied in the time domain

**NATSUKI KANDA,**[1,*] **NOBUHISA ISHII,**[1,2] **JIRO ITATANI**[1] **AND RYUSUKE MATSUNAGA**[1,3]

[1]*The Institute for Solid State Physics, The University of Tokyo, Kashiwa, Chiba 277-8581, Japan.*
[2]*Kansai Photon Science Institute, National Institutes for Quantum and Radiological Science and Technology (QST), Kyoto 619-0215, Japan.*
[3]*PRESTO, Japan Science and Technology Agency, Saitama 332-0012, Japan.*
*\*n-kanda@issp.u-tokyo.ac.jp*

**Abstract:** We report optical parametric amplification (OPA) of low-frequency infrared pulses in the intermediate region between terahertz (THz) frequency and mid-infrared (MIR), *i.e.*, from 16.9 to 44.8 THz (6.7–17.8 μm). The 255-fs laser output of the Yb:KGW regenerative amplifier is compressed to 11-fs pulses using a multi-plate broadening scheme, which generates THz-to-MIR pulses with a spectrum extending to approximately 50 THz by intra-pulse differential frequency generation (DFG) in GaSe. The THz-to-MIR pulses are further amplified using a two-stage OPA in GaSe. The temporal dynamics and photocarrier effects during OPA are characterized in the time domain. Owing to the intra-pulse DFG, the long-term phase drift of the THz-to-MIR pulses after two-stage OPA is as small as 16 mrad during a 6-h operation without any active feedback. Our scheme using the intra-pulse DFG and post-amplification proposes a new route to intense THz-to-MIR light sources with extreme phase stability.

## 1. Introduction

The development of intense ultrafast pulsed lasers has realized a novel wavelength conversion into the terahertz (THz) frequency to investigate the low-frequency electromagnetic responses in various types of matter [1–6] and mid-infrared (MIR) pulse generation by optical parametric amplification (OPA) for the higher-harmonic generation to generate coherent soft X-ray pulses [7]. More recently, the intermediate frequency between the THz (0.1–10 THz, wavelength >30 μm) frequency and the MIR (75–120 THz, or 2.5–4 μm) region, *i.e.*, from 10 to several tens of THz (4–30 μm), has garnered tremendous interest for the following reasons: (i) Compared to the THz frequency, the pulse energy can be tightly concentrated in both time and space, leading to a much higher peak field [8]. (ii) Photon energy much lower than the near-infrared (NIR) or visible range allows an intense light field to be applied in the matter without damage. (iii) The field cycle on the order of 10 fs is comparable to or even shorter than the electron scattering times and lattice motion, resulting in the emergence of nontrivial dynamics of electrons [9–11]. Furthermore, (iv) most materials have eigenmodes of lattice vibrations in this frequency range, termed as the fingerprint region, which can be strongly shaken by an intense light field for the drastic control of matter phases [12–16]. For this purpose, the time-domain waveform of the oscillating electric field plays a key role in the non-perturbative interaction regime. Therefore, intense light sources with a stable carrier-envelope phase (CEP) are in high demand.

Intense phase-locked THz-to-MIR pulses (10–70 THz) have been realized through difference frequency generation (DFG) in nonlinear crystals using two-color NIR laser pulses with the same carrier-envelope offset frequency [8, 17]. It has also been expanded for high-repetition laser systems [18] and broadband idler pulse generation [19, 20]. However, even with these novel schemes, the CEP of the output pulse can fluctuate over time owing to the long-term instability of the relative optical path length for the two different beams, leading to the requirement of an active feedback system to achieve a fluctuation of as small as 100–200 mrad [21, 22]. In this sense, intra-pulse DFG, *i.e.*, DFG within a single broadband fundamental pulse, has a significant advantage in terms of the CEP stability because of the common path for every frequency, as exemplified in optical rectification in conventional THz time-domain spectroscopy. Recently, phase stability of 56 mrad was demonstrated by the intra-pulse DFG for the output of a high-repetition 8.2-fs optical parametric chirped pulse amplifier [23]. If such a CEP-locked output is further amplified, the generation of intense THz-to-MIR pulses can be realized with excellent stability of the field waveform. Such amplification of the CEP-stabilized seed pulses has been used in optical parametric chirped-pulse amplification within the NIR region [24], and recently, within the MIR frequency of 2.7–3.8 μm (80–110 THz) [25]. To the best of our knowledge, the amplification of longer wavelengths has not been reported thus far owing to the concern of low efficiency for small photon energy.

Herein, we demonstrate the OPA of a THz-to-MIR pulse. The output of the Yb-based amplified laser is compressed and used for the intra-pulse DFG in the GaSe crystals. With an optimized phase-match condition, the generation of a frequency-tunable THz-to-MIR pulse and its amplification are also presented at between 16.9–44.8 THz (6.7–17.8 μm). The amplified pulses show excellent phase stability without active feedback. The OPA process is visualized in the time domain, and the effect of multiphoton absorption competing with the amplification is also discussed. Our demonstration of the OPA for phase-stable THz-to-MIR pulses proposes a new route to achieve a phase-stable light source for time-domain spectroscopy and sub-cycle spectroscopy.

## 2. Multi-plate pulse compression and intra-pulse DFG

We started with a Yb:KGW regenerative amplifier (PHAROS PH1-2 mJ, Light Conversion, UAB) with a pulse energy of 2 mJ, a repetition rate of 3 kHz, a pulse duration of 255 fs, and a center wavelength of 1028 nm. Because Yb-based lasers can be directly pumped by laser diodes with a high quantum efficiency, the output is stable so that the shot-by-shot pulse energy fluctuation is less than 0.04% in standard deviation with a high repetition rate. To generate a THz-to-MIR seed pulse from 15 to 50 THz, we used a GaSe crystal [26]. For intra-pulse DFG, the relatively longer pulse duration in Yb-based lasers (typically 150–300 fs) must be compressed to approximately 10 fs. Such a post-compression of Yb laser is a promising way for broadband THz-to-MIR pulse generation owing to the robustness of a diode pump system and its capability to achieve a high repetition rate [27,28]. In this study, we used the multi-plate pulse compression (MPC) method [25, 29–32] for a partial output with a pulse energy of 215 μJ, as shown in Fig. 1(a). A two-stage MPC was employed to obtain a sufficient bandwidth, similar to our previous work [25], although the original pulse duration was longer in this study. The pulse was loosely focused onto fused silica plates with a Brewster angle to avoid surface reflection loss. For a thin plate placed moderately close to the focal point while avoiding multiple filamentations, the beam profile shows a concentration in the central part by the optical Kerr effect, in which the spectrum is broadened, with another concentric outer-ring profile where the spectrum is almost unchanged. By placing the plate properly to show the concentric multi-ring shape, the divergence of the beam and the collimation with the Kerr lens effect are balanced, and the self-focusing of the beam occurs again. Thus, one can repeat the spectral broadening using multiple thin plates with an effectively large interaction length.

Four 2-mm thick plates and two 1-mm thick plates were used at the first stage of the MPC, and only the central part of the beam with a nearly uniform profile and an efficient spectral broadening was filtered with a hard aperture. The dispersion was then compensated by 6-time reflections of Gires-Tournois interferometer mirrors with negative group delay dispersion of −550 fs$^2$ (101645, Layertec GmbH). After compression in the first stage, the pulse duration was 66 fs, with a pulse energy of 73 μJ. The throughput of the first stage is ~34%, which is limited by the aperture. In the second stage, six 1-mm thick plates were used for broadening, and the pulse was compressed by 6-time reflections of a pair of chirp mirrors (Tokai Optical Co., Ltd.) with a negative group delay dispersion of −100 fs$^2$ per reflection of the mirror pair. The output pulse energy was 27 μJ, and the throughput of the second stage was ~37%. Note that the group delay dispersion of the output from multi-plates was over-compensated for the following setup, which contained the transmission in a beam splitter with a positive chirp. Several thinner fused silica plates were inserted to minimize the pulse duration at the nonlinear crystal positions for the generation and detection of the THz-to-MIR field.

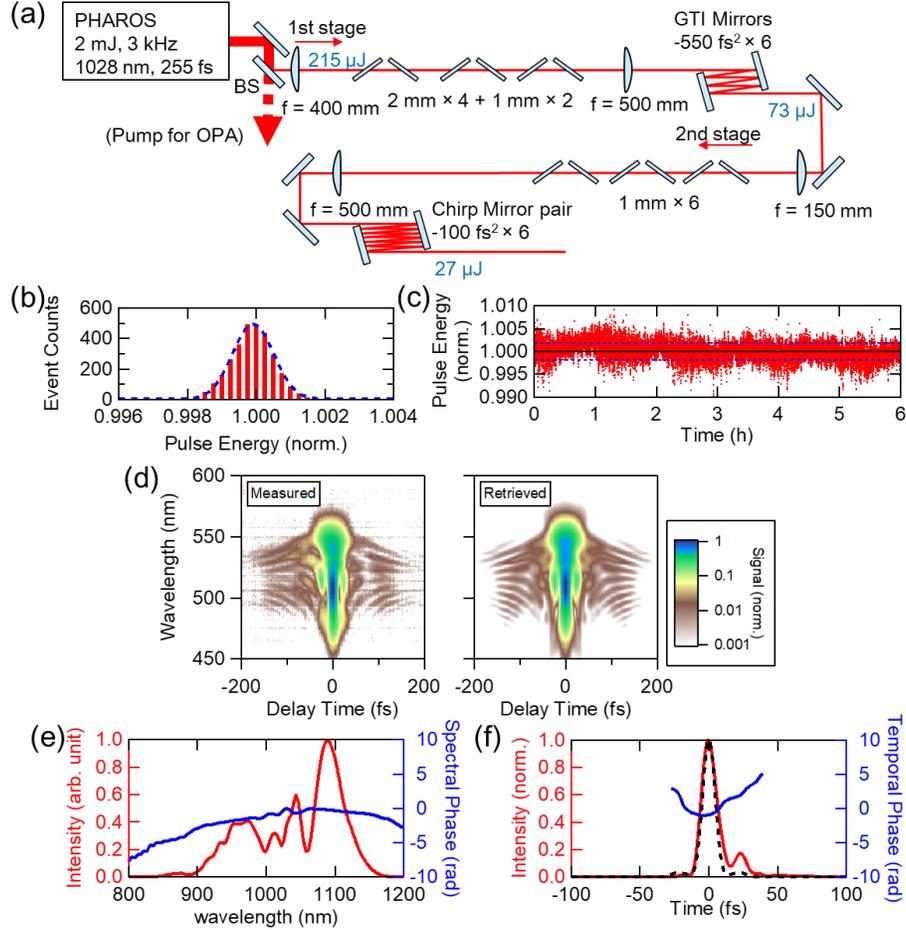

Fig. 1: Multi-plate spectral broadening and pulse compression. (a) A schematic of the setup consisted of two stages of broadening and compression. GTI represents Gires-Tournois interferometer mirrors. (b) Histogram of the output pulse energy of 3,000 pulses measured using shot-by-shot detection (red). Normal distribution function with a standard deviation of 0.06% is also shown (blue). (c) Long-term stability of output power of compressed pulses in 6 hours (red dots). The standard deviation is 0.18% (blue dashed). (d) Measured and retrieved SHG-FROG traces of the output of the second stage compression. The FROG error is 0.9% over a 512 × 512 retrieval grid. (e) Retrieved intensity spectrum (red) and spectral phase (blue). (f) Time-domain intensity envelope of the retrieved pulse (red) and temporal phase (blue). Fourier-transform-limited pulse is also shown by a black dashed line.

Figure 1(b) shows the distribution of the output pulse energy measured using a photodiode for each pulse. Remarkably, the shot-by-shot pulse energy fluctuation was as small as 0.06% of the standard deviation even after nonlinear spectral broadening and compression, which corroborates the stability of this compression method available for a variety of optical systems. Long-term stability was also measured using a thermal detector (PM10, Coherent, Inc.), and the fluctuation was 0.18% in standard deviation for 6 h, as shown in Fig. 1(c). The compressed pulse was characterized using a second-harmonic generation-based frequency-resolved optical gating (SHG-FROG) method. The measured and retrieved FROG traces are presented in Fig. 1(d). Figure 1(e) illustrates the retrieved intensity and phase spectra, showing a sufficiently broad spectrum to generate pulses of several tens of THz with an intra-pulse DFG. Figure 1(f) shows the time-domain waveform of the pulse envelope in the temporal phase. The compressed pulse duration was as short as 11 fs, which was close to a Fourier-transform limited pulse.

We applied this stable compressed pulse to the generation and time-domain observations of THz-to-MIR pulses. A broadband THz-to-MIR pulse was generated using an intra-pulse DFG in a *c*-axis-cleaved GaSe crystal with a thickness of 14 μm. Another thin GaSe crystal of 3.5 μm in thickness is used for the detection of the THz-to-MIR pulse using the electro-optic sampling (EOS) method with a gate pulse. By using these thin crystals, a broadband pulse can be generated and detected without being restricted by the phase-matching condition [33–35]. The emitted THz-to-MIR beam was collimated and focused using off-axis parabolic mirrors to the detection crystal. Figures 2(a) and 2(b) show the time-domain waveform of the electric field and the power spectrum, respectively, indicating a short-pulsed waveform and a broad power spectrum of up to 50 THz.

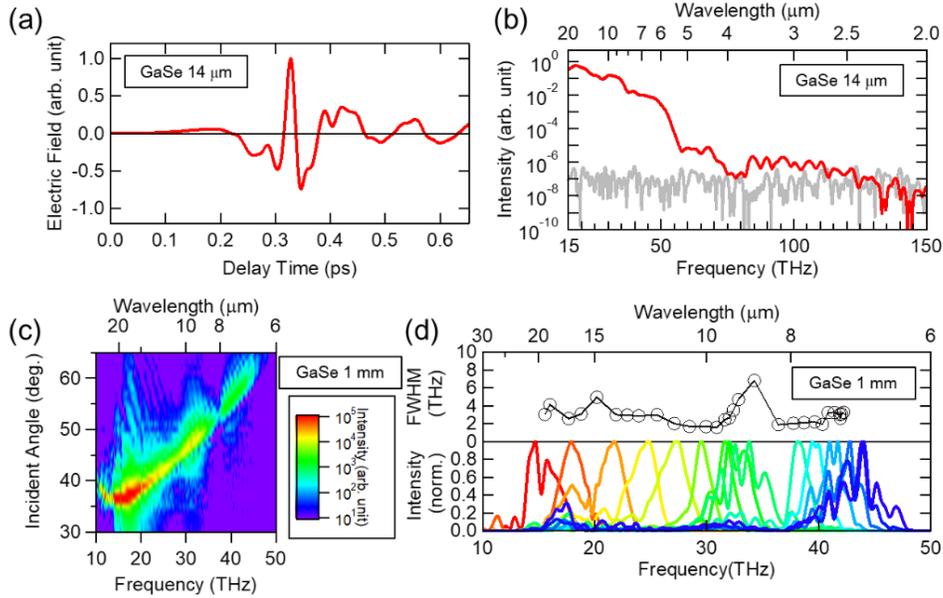

Fig. 2: (a) Time-domain waveform of a broadband THz-to-MIR pulse generated by intra-pulse DFG with a 14-μm thin GaSe crystal. (b) Power spectrum of the waveform in (a). The gray curve shows a noise floor measured without THz-to-MIR pulses. (c) Two-dimensional plot of the power spectra for the narrowband THz-to-MIR pulse by intra-pulse DFG with a 1-mm thick GaSe crystal in logarithmic scale as functions of the incident angle and frequency. (d) Normalized power spectra (lower) and FWHM (upper).

We also used another 1-mm thick GaSe crystal for the emitter to increase the THz-to-MIR pulse intensity for a specific frequency. Here, we employed the type-II phase matching, *i.e.*, $\omega_1(e) - \omega_2(o) = \Omega(e)$, where $\omega_1$ and $\omega_2$ are the frequencies in the NIR contained within the fundamental pulse, $\Omega$ is the DFG frequency, and "o" and "e" represent ordinary and extraordinary polarized lights, respectively. By changing the incident angle, the DFG frequency $\Omega$ can be tuned according to the phase-matching condition [18, 26, 33]. The polarization of the incoming laser beam was rotated by 45° in order to obtain both ordinary and extraordinary projections in the reference frame of the crystal. Figure 2(c) shows a two-dimensional plot of the power spectra of the generated THz-to-MIR pulses with different incident angles. The peak draws an arc from approximately 16 to 45 THz, which is consistent with the previous calculation of the phase match condition for GaSe [18]. Figure 2(d) shows the normalized spectra and bandwidth. The distortions of the spectra and bandwidth at approximately 20 and 37 THz can be ascribed to the absorption of $CO_2$ in air and the absorption loss of the dielectric layers of protection-coated metallic mirrors, respectively.

### 3. OPA in THz-to-MIR region

Herein, we demonstrate the amplification of the phase-stable THz-to-MIR pulse using the OPA process, which is a key technology for generating intense light sources. The phase-stable seed pulse using an intra-pulse DFG is essential for realizing phase-stable intense pulses through the OPA process. Figure 3(a) shows a schematic of the OPA. The pump delay was scanned to tune the temporal overlap of the seed and pump pulses, and the gate delay was also scanned for the characterization of the output pulse in the time domain.

The THz-to-MIR pulse generated by the intra-pulse DFG with a 1-mm thick GaSe crystal shown in Fig. 2(d) was used as a seed, which was focused onto another 2-mm thick GaSe crystal for the OPA. The pump pulse with a pulse energy of 277 μJ at 1028 nm was separated before the MPC and focused on an OPA crystal. The $1/e^2$ beam diameter of the pump was 1.82 mm at the crystal position, which corresponds to a fluence of 21.3 mJ/cm$^2$ and a peak intensity of 78.5 GW/cm$^2$. The phase match condition for the OPA is almost the same as that of the DFG because only the roles of $\omega_2$ and $\Omega$ are replaced. In the type-II phase-matching condition, both the seed and pump pulses have *p*- polarizations on the OPA crystal, in which the reflection loss can be suppressed compared with the type-I phase matching. Finally, the transmitted field is detected using the EOS with the gate pulse.

Figure 3(b) shows the output transmitted field waveforms at pump delays of 0 and −10 ps. The corresponding power spectra with a peak frequency of 17.7 THz are shown in Fig. 3(c). The temporal durations of the OPA output and seed were 437 and 572 fs, respectively. The output pulse energy was 17 nJ, which was measured with a thermal detector (3A-P-THz, Ophir Optronics Ltd.) The energy conversion efficiency from NIR pump to THz-to-MIR signal was estimated as $6.1 \times 10^{-5}$ in this first stage OPA. At a pump delay of −10 ps, the seed pulse simply passes through the crystal long before the pump arrives (seed-first limit). In contrast, at a pump delay of 0 ps, the pump and seed pulses temporally overlap, resulting in an enhancement of the seed, as clearly observed. This is the first observation of the OPA as an electric field in the time domain to the best of our knowledge. Figure 3(d) shows a two-dimensional plot of the output waveform as a function of the gate and pump delays. The dashed line indicates the timing of the pump pulse entering the OPA crystal, around which the amplification is clearly observed. The upper-right and lower-left sides of the dashed line correspond to the pump- and seed-first conditions, respectively.

The intensity and phase shift at the peak frequency as a function of the pump delay are also shown by the red markers in Figs. 3(e) and 3(f). The intensity was estimated by integrating the power spectrum and normalized with that at the seed-first limit. The phase shift is defined as the phase difference from the phase at the peak frequency in the seed-first limit. The output intensity was amplified by approximately 100 at around the zero pump delay and then drastically decreased after the pump pulse passed through the crystal before the seed (red markers in Fig. 3(e)). This can be attributed to the absorption of free carriers excited by the pump with two-photon absorption. For comparison, we also conducted a similar measurement with a pump of the ordinary polarization, where the phase-matching condition is not satisfied, as plotted by blue markers in Figs. 3(e) and 3(f). In this case, amplification did not occur, and only the depletion of the intensity was observed. The difference in the signal level at the pump-first limit in Fig. 3(e) is ascribed to the polarization-dependent reflection loss of the pump. Figure 3(e) shows that the intensity of the extraordinary pump starts to increase at around −0.4 ps and reaches the peak at zero pump delay, while the suppression of transmission for the ordinary pump occurs a few hundred femtoseconds later. One might think that the difference in the timing in Fig. 3(e) could be attributed to the different refractive indices for the different polarizations. However, Fig. 3(f) shows that the phase shifts start at almost the same pump delay for both polarizations, indicating that the influence of the pump occurs at almost the same delay for both polarizations. The result indicates that the amplification can occur prior to the free-carrier absorption, which could be the main reason why OPA is possible against the two-photon absorption in this work. We also found that the waveform of the amplified pulse in Fig. 3(b) shows a noticeable asymmetry and rapid decay of the envelope for the latter half. This can be explained by the free-carrier absorption, which mainly influences the latter half of the amplification.

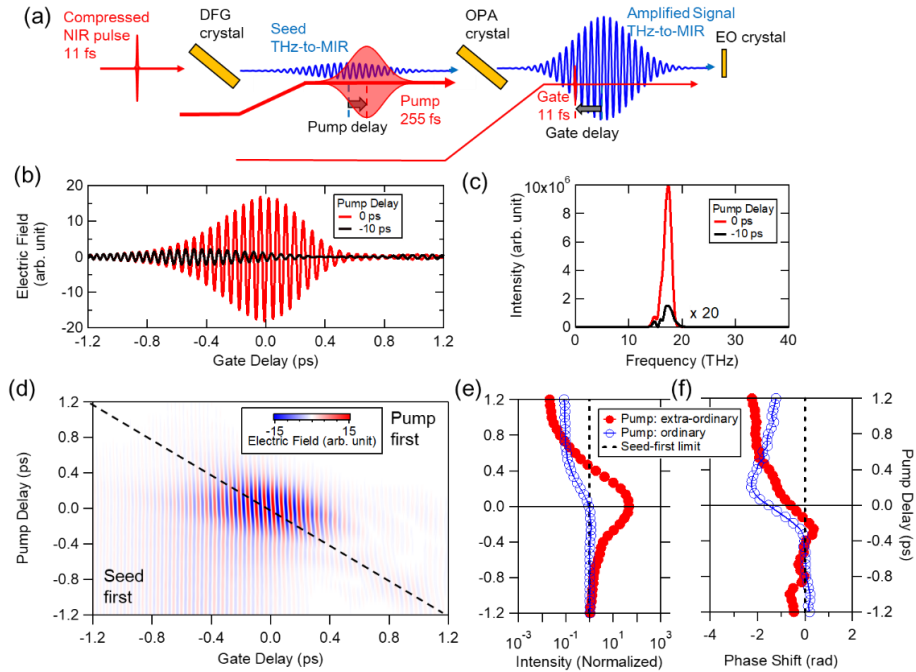

Fig. 3: (a) A schematic of time-resolved measurement of OPA dynamics. (b), (c) Time-domain waveforms and the power spectra of the transmitted THz-to-MIR pulse at zero pump delay (red) and the seed-first limit (pump delay of −10 ps) (black). Note that the spectrum at the seed-first limit is magnified by 20 for visibility. (d) Two-dimensional plot of the time-domain waveforms for the amplified pulse as a function of the gate delay (horizontal) and pump delay (vertical). (e) The pump-delay dependence of the intensity of the data in (d) is shown by the red markers. For comparison, the data for the ordinary polarization pump (phase-mismatched condition) is also shown by blue markers. The black dashed line shows the signal level at the seed-first limit. (f) The corresponding results for the phase shift.

## 4. Second-stage OPA

Next, we demonstrate the second-stage OPA for further amplification using residual pump power. Another 2-mm thick GaSe crystal was placed similarly to the first-stage OPA, and a part of the divided pump beam with a pulse energy of 567 µJ was used for the second stage. The beam diameter was 2.34 mm, which corresponds to a fluence of 26.4 mJ/cm$^2$ and a peak intensity of 97.1 GW/cm$^2$. The average power of the amplified beam was measured with a thermal detector, and the electric-field waveform was obtained using the EOS. To discuss the scalability of our OPA, the pump and seed power dependences were measured. These experiments were conducted with a peak frequency of 18.1 THz.

Figure 4(a) shows the pump-power dependence of the output power after the second-stage OPA. The dashed line in Fig. 4(a) shows a linear increase at a lower pump power. The output power increases super-linearly at the lower pump and becomes closer to linear dependence at a higher pump. At the maximum pump, an average output of 800 µW (pulse energy of 267 nJ) was obtained. The absence of super-linearity in the higher pump power regime can be attributed to two-photon absorption. Nevertheless, the OPA output was still linear, and the saturation of the output power was not identified in this pump regime. The seed power dependence is shown in Fig. 4(b). To regulate the seed power in the second stage (*i.e.*, the output of the first stage), the efficiency of the intra-pulse DFG was controlled by tuning the polarization of the NIR pulse before the intra-pulse DFG crystal. The seed power dependence was almost linear, and saturation was not observed. These results suggest the possibility of further amplification at multi-stage OPA.

Figure 4(c) shows the waveform at the maximum pump power with a pulse duration of 469 fs in FWHM, which was 1.7 times larger than the Fourier limit. The spot size was measured using a bolometer focal-plane-array camera for THz radiation (IRV-T0830, NEC Co.) [36], as shown in Fig. 4(d). The profile was well fitted with Gaussian functions, and the beam widths (full width at 1/e$^2$ of maximum) were evaluated as 200 and 113 µm for the horizontal and vertical directions, respectively. From these evaluations, we estimated the peak electric field to be 2.07 MV/cm. The spot size was still several times larger than the diffraction limit because of the aberration of the off-axis parabolic mirrors. A higher peak field can be achieved by improving the focusing conditions.

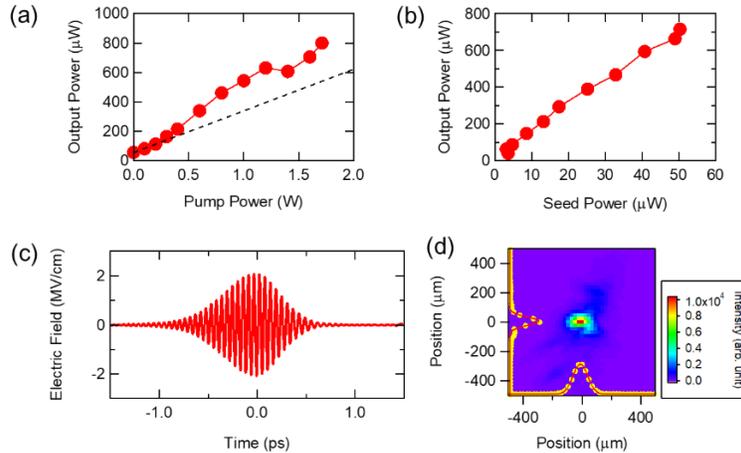

Fig. 4: Results of second-stage OPA at 18.1 THz: (a) Pump power dependence with a fixed seed power of 50 µW. The dashed line shows linearity for weak pump power. (b) Seed power dependence with a fixed pump power of 1.7 W. (c) Waveform of an electric field at the maximum pump power. (d) Focal point image obtained using a THz camera with a pixel size of 23.5 µm.

We also demonstrate the frequency tuning of the OPA. As shown in the intra-pulse DFG in Fig. 2(b), the peak frequency of the amplified signal can be tuned by changing the incident angle to the OPA crystal. The lower image in Fig. 5(a) shows the normalized spectra of the amplified pulse for different incident angles of 36 to 56° for tuning the peak frequency from 16.9 to 44.8 THz. For the measurements of power spectra above 29 THz, four Si wafers were inserted as attenuators with their reflection loss to obtain an EOS signal from only the Pockels effect. The bandwidth of the spectra is 0.7-1.2 THz as shown in the middle of Fig. 5(a). The output power is shown in the upper part of Fig. 5(a). The maximum output power was 2 mW at 33.3 THz, which corresponds to a conversion efficiency of $1.2 \times 10^{-3}$. The large dip at approximately 37 THz is ascribed to the reflection loss of the protection-coated gold mirrors, as shown on the right axis in Fig. 5(a), which can be avoided by using bare gold mirrors.

Finally, we evaluated the long-term stability of the amplified THz-to-MIR pulse after the two OPA stages. Figure 5(b) shows a two-dimensional plot of the time-domain waveform repeatedly measured during a 6-h period. The phase shift of each measurement is shown on the right side. The fluctuation of the phase was as small as 16 mrad in the standard deviation without any active feedback. Such high phase stability in the amplified pulse without active feedback is realized owing to the intra-pulse DFG method and is several times or an order of magnitude better than the inter-pulse DFG in previous studies, even with active feedback.

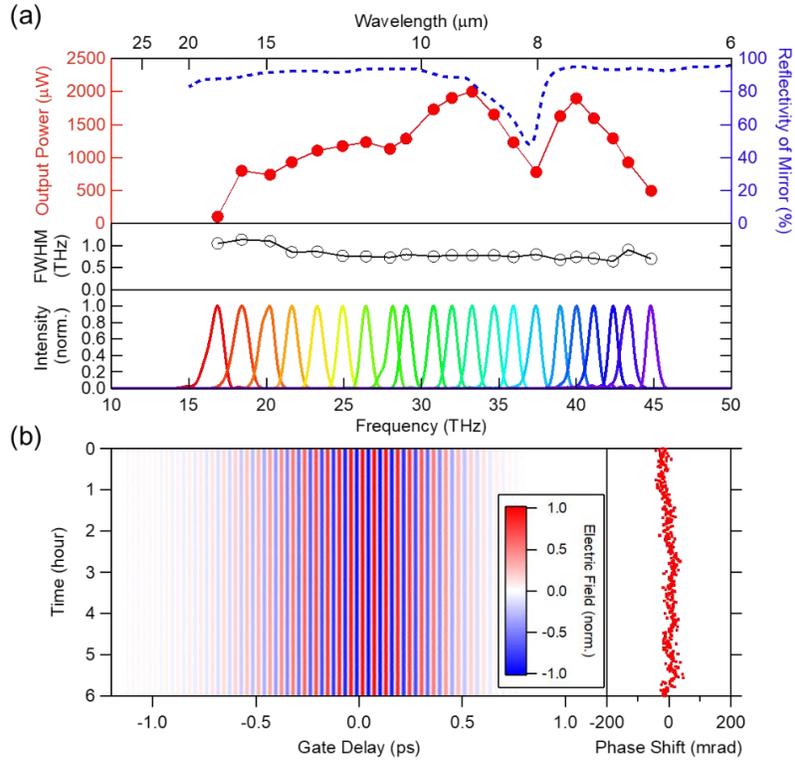

Fig. 5: (a) Tuning of OPA by changing the incident angle. Peak-frequency dependence of the output intensity (upper), FWHM of the power spectra (middle), and normalized power spectra from incident angles of 36 to 56° (lower). The blue curve on the right axis shows the reflectivity of the protection-coated mirrors used in this system. (b) Stability of the waveform of the amplified signal for a 6-h period. A phase shift is also shown on the right side.

## 5. Conclusion

We report the optical parametric amplification of phase-stable THz-to-MIR pulses in the unexplored low-frequency region of 17 to 45 THz (6.7-18 μm). After the two-stage OPA, the field strength reached more than 2 MV/cm, and the non-saturating behaviors in the pump and seed power dependences suggest possible further amplification. Notably, owing to the seed pulse generation by the intra-pulse DFG, the phase of the amplified signal was quite stable with a fluctuation of 16 mrad in standard deviation during the 6-h operation without any active stabilization, even though our demonstration contains two-stage pulse compression of NIR pulses and two-stage amplification of THz-to-MIR pulses. Our method opens a new route for obtaining high-intensity and phase-stable THz-to-MIR light sources for infrared science. Other nonlinear crystals such as $LiGaS_2$ or $LiGaSe_2$ are also promising for high-frequency regions of up to 60 THz [18, 19, 37].

Our time-domain observation of the OPA process also provides insights into OPA under the condition that multiphoton absorption exists. Although GaSe crystals and a Yb-laser are a realistic choice for amplification in this frequency region at present, future development of MIR pump lasers and/or alternative nonlinear crystals may realize much more efficient amplification suppressing multiphoton absorption.

## Funding

This work was supported in part by JSPS KAKENHI (Grant Nos. JP19K15462, JP18H01843, JP19H02623, and JP18H05250), and JST PRESTO (Grant No. JPMJPR16PA), Murata Foundation, and MEXT Quantum Leap Flagship Program (MEXT Q-LEAP), Grant No. JPMXS0118068681.

## Acknowledgments

We thank Y. Murotani for fruitful discussions and technical support.

## Disclosures

The authors declare no conflicts of interest.